\documentclass[preprint,aps,amsmath]{revtex4}
\usepackage{graphicx}
\usepackage{subfigure}
\usepackage{units}
\usepackage{color}
\begin{document}

\title{Analysis of the temperature influence on Langmuir probe measurements on
  the basis of gyrofluid simulations} 

\author{Felix P. Gennrich and Alexander Kendl}

\address{Institute for Ion Physics and Applied Physics, Association
  Euratom-\"{O}AW, University of Innsbruck, Technikerstra{\ss}e 25, A-6020
  Innsbruck, Austria \vspace{1.5cm} } 

\begin{abstract}
\vspace{0.5cm} 
The influence of the temperature and its fluctuations on the ion saturation
current and the floating potential, which are typical quantities measured by
Langmuir probes in the turbulent edge region of fusion plasmas, is analysed by
global nonlinear gyrofluid simulations for two exemplary parameter
regimes. The numerical simulation facilitates a direct access to densities,
temperatures and the plasma potential at different radial positions around the
separatrix. This allows a comparison between raw data and the calculated ion
saturation current and floating potential within the simulation. Calculations
of the fluctuation-induced radial particle flux and its statistical properties
reveal significant differences to the actual values at all radial positions of
the simulation domain, if the floating potential and the temperature averaged
density inferred from the ion saturation current is used. 

\vspace{6cm}
{\sl This is the preprint version of a manuscript submitted to Plasma Physics
  and Controlled Fusion.}

\end{abstract} 

\maketitle

\section{Introduction}

In the edge and scrape-off layer regions of magnetically confined plasmas the
fluctuating plasma density ${n = n_{\rm{e}} \simeq n_{\rm{i}}}$, the plasma
potential $\Phi$ and the radial particle flux $\Gamma_{\rm{r}}$ are usually
inferred from Langmuir probe measurements. 
However, the quantities measured by conventional cold Langmuir probes are the
ion saturation current $I_{\rm{is}}$ and the floating potential
$V_{\rm{fl}}$. Following the elementary Langmuir probe theory, these are
related to the density and the plasma potential by expressions involving the
electron and ion temperatures $T_{\rm{e}}$ and $T_{\rm{i}}$
(\cite{Schrittwieser2002}--\cite{Stangeby1990}): 
\begin{eqnarray}
 I_{\rm{is}} & = & A_{\rm{i}} e n \sqrt{\frac{k_{\rm{B}}(T_{\rm{e}}+T_{\rm{i}})}{m_{\rm{i}}}} 
\label{eq:isat} \\
 V_{\rm{fl}} & = & \Phi - \left( \frac{k_{\rm{B}} T_{\rm{e}}}{e} \right) \ln
\left( \frac{I_{\rm{es}}}{I_{\rm{is}}} \right) = \Phi - \left(
\frac{k_{\rm{B}} T_{\rm{e}}}{e} \right) \ln \left(
\frac{A_{\rm{e}}}{A_{\rm{i}}} \sqrt{\frac{T_{\rm{e}}}{T_{\rm{e}}+T_{\rm{i}}}}
\sqrt{\frac{m_{\rm{i}}}{2 \pi m_{\rm{e}}}} \right).
\label{eq:phifloat}
\end{eqnarray}

A Maxwellian electron velocity distribution is assumed, secondary
electron emission from the probe is neglected and the electron saturation
current $I_{\rm{es}}$ is given by $I_{\rm{es}} = A_{\rm{e}} e n (1/4) \sqrt{(8
  k_{\rm{B}} T_{\rm{e}}) / (\pi m_{\rm{e}})}$, using the random thermal
current density. $A_{\rm{e}}$ and $A_{\rm{i}}$ specify the probe collecting
areas for electrons and ions, respectively. Depending on the magnetic field
strength, these areas can be differing for the two species, as stated in
\cite{Schrittwieser2002}. At any rate, to determine density and plasma
potential from the measured quantities, electron and ion temperatures have to
be taken into account, although $T_{\rm{i}}$ is frequently assumed to be equal
to $T_{\rm{e}}$. This is mainly due to the fact that in the edge region of
fusion devices there is often no data available for the ion temperature. 

The measurement of electron temperature fluctuations by means of classical
Langmuir probes requires a sweeping of a preferably complete probe
characteristic. This results in a lower time resolution of the respective time
series compared to data acquired by floating or negatively biased
$I_{\rm{is}}$ probe pins, although the method has undergone further
development towards fast sweeping probes
(\cite{Mueller2010}-\cite{Hidalgo1992}). Alternatively, triple probes
(\cite{Chen1965}, \cite{Xu2007}) or the harmonics technique
(\cite{Boedo1999}-\cite{Rudakov2001}) can be used to measure fluctuations of
$T_{\rm{e}}$. In addition, more sophisticated probes such as \textit{emissive
  probes} (\cite{Schrittwieser2002}, \cite{Schrittwieser2008}), in contrast to
conventional \textit{cold probes}, or \textit{ball-pen probes}
(\cite{Horacek2010}, \cite{Adamek2010}) have been developed. These kind of
probes are aimed at measuring the plasma potential directly and a combination
of cold probes and ball-pen or emissive probes also allows a derivation of the
electron temperature \cite{Adamek2002}. Nevertheless most probe measurements
in the edge of large fusion devices are still based upon data measured by
classical Langmuir probes. For estimations of the radial particle flux
$\Gamma_{\rm{r}} = \tilde{n} \tilde{v_{\rm{r}}}$, the radial velocity is
commonly calculated from gradients of the floating potential instead of the
plasma potential \cite{Kirk2011} and the density is calculated from
$I_{\rm{is}}$ using only average values for the temperatures. 

The aim of this paper is to use numerical gyrofluid simulations to compare
time series of density $n$, plasma potential $\Phi$ and temperatures
$T_{\rm{e}}$, $T_{\rm{i}}$ with simulated values of ion saturation current
$I_{\rm{is}}$ and floating potential $V_{\rm{fl}}$. The difference between the
two corresponding datasets of $\Phi$ and $V_{\rm{fl}}$ is to some extent
comparable to the difference between emissive and conventional probe
measurements, although the potential measured by the former still deviates
from the actual plasma potential by a certain, albeit smaller
temperature-dependent factor \cite{Schrittwieser2008}. 

For the simulations the nonlinear three-dimensional electromagnetic gyrofluid
turbulence code \textit{GEMR} has been used, which comprises a six-moment
gyrofluid model for electrons and ions in a circular toroidal geometry and features energetic
consistency (\cite{Scott2006}-\cite{Ribeiro2008}). The coordinate system in
use consists of a flux surface label ($x$) defining the radial position, a
field line label within the flux surface ($y$) and a position along the field
line ($s$) \cite{Scott2001}. For diagnostic purposes, the code delivers time
series of fluctuating electron and ion densities, plasma potential,
temperatures and parallel velocities, amongst others. Thus, the knowledge of
these quantities including electron and ion temperatures allows the
calculation of ion saturation current and floating potential. So the
significance of temperature fluctuations with regard to density and potential
measurements can be investigated. Moreover, the analysis can be performed at
different simulated probe positions in the radial computation domain, which in
our nominal case is $(r/a = 1 \pm 0.06)$, with $a$ being the minor radius. 

Typical ASDEX Upgrade edge values have been chosen as input parameters for the
simulation. The model can be regarded as global in the sense that there is a
global variation of profiles, although the parameters are constant
\cite{Scott2006}. Although no turbulence code can as yet self-consistently
achieve an H--mode, magneto-hydrodynamic ideal ballooning modes (IBMs), which
are commonly assumed to cause edge localised modes of type I, can be simulated
by incorporating experimental H-mode density and temperature pedestal profiles
\textit{n}(\textit{r}) and \textit{T}(\textit{r}) as initial state
\cite{Kendl2010}. Ideal ballooning (ELM-like) blowouts in experiment and
simulation are always connected with large fluctuations in density,
temperature and plasma potential, so a comprehensive analysis of the
temperature influence on floating potential and ion saturation current in this
case is a reasonable addition to investigations of simulations in saturated
L-mode state. 

In the following sections the details of the comparison between plasma density
and ion saturation current, plasma potential and floating potential and
characteristics of the particle flux in L-mode situation (section 2) as well
as IBM blowout situation (section 3) are presented. Possible reasons for the
varying discrepancy between the actual quantities and the simulated Langmuir
probe measurements are also addressed and the role of temperature fluctuations
is discussed.

\section{Saturated L-mode situation}

The first situation corresponds to an operation in saturated L-mode. That
means, the electron dynamical plasma beta $\beta_{\rm{e}} = (\mu_{0}
p_{\rm{e}})/B^{2}$ is chosen low enough not to be ideal ballooning unstable
($\beta_{\rm{e}} \approx 9.4 \cdot 10^{-5}$) and electron and ion heat sources
as well as density sources are set to a moderate level. The background
mid-pedestal parameters for this case are $T_{\rm{e}} = 150$ eV and
$T_{\rm{i}} = 180$ eV for the temperatures, $n_{\rm{e}} = n_{\rm{i}} = 1.25
\cdot 10^{19}$ m$^{-3}$ for the densities, $B = 2.0$ T for the background
magnetic field, $L_{\rm{T}} = L_{\perp} = 3.0$ cm for the perpendicular
temperature gradient length and $L_{\rm{n}} = 6.0$ cm for the density gradient
length. For the ion mass, the deuterium mass $m_{\rm{D}} = 3670 m_{\rm{e}}$ is
used. Major torus radius and aspect ratio (major divided by minor radius)
conform to ASDEX Upgrade values with $R = 1.65$ m and $R/a = 3.3$, while a
circular flux-surface geometry is used. 

The radial domain, whose direction is
defined as the $x$-direction in the code, is divided into $n_{x} = 64$ grid
points. These are not equally spaced in terms of radius but in terms of
volume, which for the present purpose makes a slight but not decisive
difference. The $y$-direction is divided into $n_{y} = 512$ grid points, the
$s$-direction into $n_{s} = 16$ grid points and the averaged grid size
perpendicular to the magnetic field is $1.06 \rho_{\rm{s}} \times 1.39
\rho_{\rm{s}}$ (with $\rho_{\rm{s}} \approx 0.88$ mm). The analysed L-mode
time series have a length of $11000$ data points and a temporal resolution of
$\unit[0.0826]{\mu s}$ (total duration $\unit[0.908]{ms}$). 

Especially in the outer scrape-off layer region of the radial domain ($x \sim
33 - 64$) the numerical (Arakawa)--scheme occasionally delivers unphysical
negative density and temperature values in the presence of steep propagating
gradients, arising from Gibbs oscillations. Therefore we added the absolute
value of the largest negative value multiplied by an offset factor
$\lambda_{\rm{os}}$ to all values of the time series at all positions,
e.g. for the density, $n_{\rm{i},\mathit{x}}$ = $\bar{n}_{\rm{i},\mathit{x}} +
\lambda_{\rm{os}} |\rm{Min}({\bar{n}_{\rm{i}}})|$, where
$\bar{n}_{\rm{i},\mathit{x}}$ is the uncorrected time series at $x$ and the
minimum of all values in the radial domain is used. As offset factor,
$\lambda_{\rm{os}} = 1.01$ has been chosen. 

\begin{figure}
 \centering
 \subfigure[]{\label{fig:lmode:raw:tepl}\includegraphics[width=7.5cm]{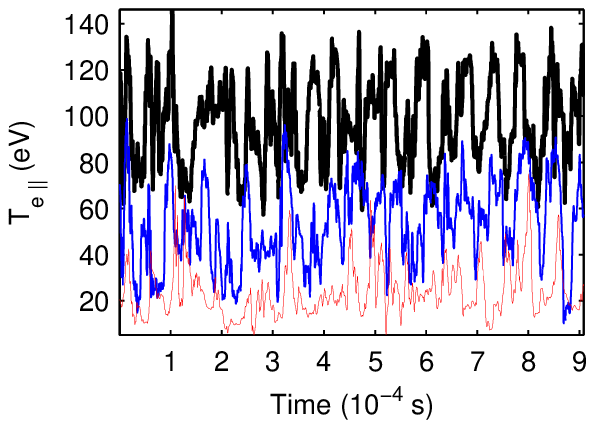}}\hfill
 \subfigure[]{\label{fig:lmode:raw:all}\includegraphics[width=7.5cm]{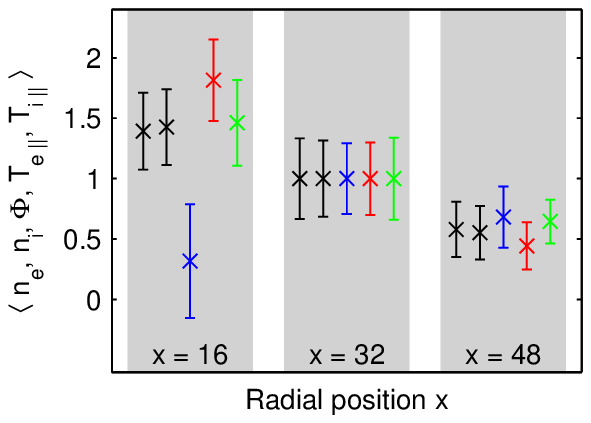}}
\caption{\sl (a) Time series of the electron temperature $T_{\rm{e}
    \parallel}$. (b) Mean values and standard deviations of electron and ion
  density $n_{\rm{e}}$ and $n_{\rm{i}}$ (black), plasma potential $\Phi$
  (blue), electron temperature $T_{\rm{e} \parallel}$ (red) and ion
  temperature $T_{\rm{i} \parallel}$ (green), normalised to the respective
  mean at $x = 32$, for the L-mode case at different radial positions. The
  simulation grid point $x = 16$ (black bold line in (a)) corresponds to the
  radial distance of $\sim -14.8$ mm $\approx -16.7 \rho_{\rm{s}}$, $x = 32$
  (blue line in (a)) to $\sim 0.4$ mm $\approx 0.5 \rho_{\rm{s}}$ and $x = 48$
  (red thin line in (a)) to $\sim 15.2$ mm $\approx 17.2 \rho_{\rm{s}}$,
  measured from the separatrix} 
\label{fig:lmode:raw}
\end{figure}

The simulation run exhibits an ion temperature gradient (ITG) driven
interchange overshoot quite at the beginning and then saturates rather quickly
to an L-mode-like state. For the analysis and the plots presented in this
section only the well saturated part after the initial transient has been
taken into account. Fig.~\ref{fig:lmode:raw} exemplary shows the time series of
$T_{\rm{e} \parallel}$ as well as mean values and standard deviations of
$n_{\rm{e}}$ and $n_{\rm{i}}$, $\Phi$, $T_{\rm{e} \parallel}$ and $T_{\rm{i}
  \parallel}$ at different radial positions (inside, near and outside the
separatrix). 

Both here and in the following parts of the L-mode section, plots
showing the temporal evolution of a quantity are based on data taken near the
outboard midplane, whereas plots of time averaged data are composed from the
individual results of different toroidal positions to provide better
statistics. The radial position outside the separatrix ($\sim 1.5$ cm, red
thin lines) corresponds approximately to the region of Langmuir probe
measurements in ASDEX Upgrade, which are typically positioned few centimeters
outside the last closed flux surface. A separate treatment of parallel and
perpendicular temperatures ($T_{\rm{e} \parallel}$, $T_{\rm{i} \parallel}$ and
$T_{\rm{e} \perp}$, $T_{\rm{i} \perp}$) arises from the construction of
moments in the gyrofluid equations \cite{Beer1996}. However, in the present
case there is no major difference between both components and only time series
of the parallel temperatures have been used to calculate the quantities of the
synthetic probe.

\subsection{Plasma density vs. ion saturation current}
\label{sec:lmode:nvsisat}

The ion saturation current $I_{\rm{is}}$ has been calculated from ion density
and the temperatures of electrons and ions according to eq.~\ref{eq:isat} (with
$A_{\rm{e}} = A_{\rm{i}}$ defining the cross section of the probe). As the
characteristics of the $n_{\rm{i}}$, $T_{\rm{e} \parallel}$ and $T_{\rm{i}
  \parallel}$ time series are rather similar, there are no striking
differences between the ion saturation current, which is essentially a product
between them, with the sum of temperatures appearing as a square root, and the
underlying quantities and it still shows a qualitatively comparable temporal
evolution. 

\begin{figure}
 \centering
 \subfigure[]{\label{fig:lmode:isat:3}\includegraphics[width=7.5cm]{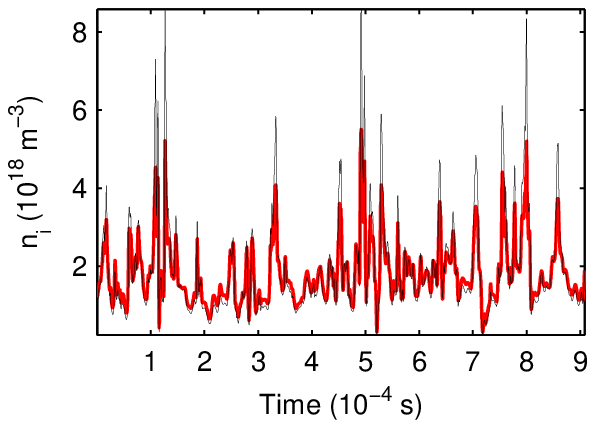}}\hfill
 \subfigure[]{\label{fig:lmode:isat:3part}\includegraphics[width=7.5cm]{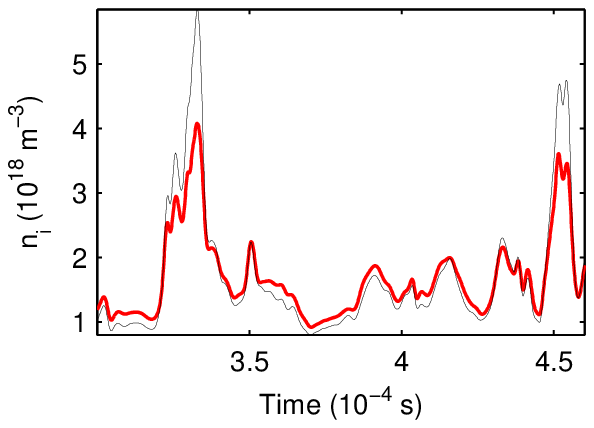}}
\caption{\sl (a) L-Mode time series of code output $n_{\rm{i}}$ (red bold line)
  and $n_{\rm{i}}^{\rm{avg}}$ (black thin line) calculated from $I_{\rm{is}}$
  using averaged values for $T_{\rm{e} \parallel}$ and $T_{\rm{i}
    \parallel}$. The data is plotted for a radial position $\sim 15.2$ mm
  $\approx 17.2 \rho_{\rm{s}}$ outside the separatrix. (b) Detailed evolution
  of a $\unit[160]{\mu s}$ time frame} 
\label{fig:lmode:isat}
\end{figure}

For evaluations of the radial particle flux from experimentally
measured probe data it is necessary to know, amongst others, the particle
density. Although the conversion can easily be done using eq.~\ref{eq:isat}, in
many experimental cases only average temperature values are available
(e. g. evaluated by sweeping the Langmuir probe). Hence it makes sense to
recalculate $n_{\rm{i}}$ from $I_{\rm{is}}$ using averaged values of the
simulated $T_{\rm{e} \parallel}$ and $T_{\rm{i} \parallel}$ time series for
each $x$-position in the radial domain and to compare the results with the
precise $n_{\rm{i}}$ time series. Both signals are plotted in
fig.~\ref{fig:lmode:isat} for a position within the SOL. Inside and near the
separatrix similar results have been obtained. The difference between
$n_{\rm{i}}^{\rm{avg}}$ and $n_{\rm{i}}$, given by 
\begin{equation}
n_{\rm{i}}^{\rm{avg}} - n_{\rm{i}} = \frac{I_{\rm{is}}
  \sqrt{m_{\rm{i}}}}{A_{\rm{i}} e} \left(\frac{1}{\sqrt{k_{\rm{B}}
    (T_{\rm{e}}^{\rm{avg}} + T_{\rm{i}}^{\rm{avg}})}} -
\frac{1}{\sqrt{k_{\rm{B}} (T_{\rm{e}} + T_{\rm{i}})}}\right), 
\label{eq:nidiff}
\end{equation}
 is particularly large at points of maxima and minima, but rather
small elsewhere. As $n_{\rm{i}}^{\rm{avg}} = n_{\rm{i}} \sqrt{(T_{\rm{e}} +
  T_{\rm{i}})/(T_{\rm{e}}^{\rm{avg}} + T_{\rm{i}}^{\rm{avg}})}$, the ratio of
the actual temperature values to their average values is the decisive
factor. The average temperatures are smaller than the maxima, which results in
larger values of $n_{\rm{i}}^{\rm{avg}}$ at these positions (the temperatures
appear in the denominator in eq.~\ref{eq:nidiff}). For the minima, the same
statement holds vice versa. 

\subsection{Plasma potential vs. floating potential}
\label{sec:lmode:phivsvfl}

\begin{figure}
 \centering
 \subfigure[]{\label{fig:lmode:phi:3}\includegraphics[width=7.5cm]{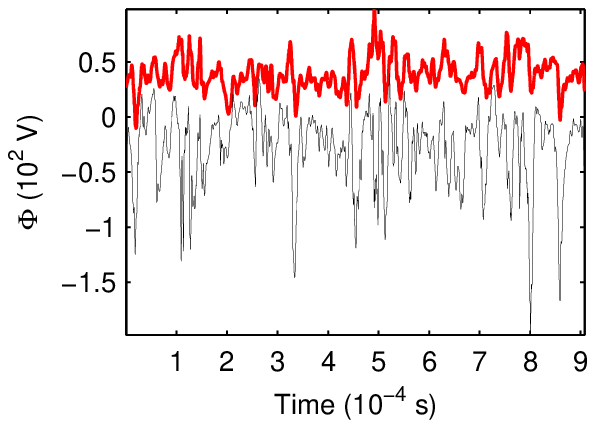}}\hfill
 \subfigure[]{\label{fig:lmode:phi:3part}\includegraphics[width=7.5cm]{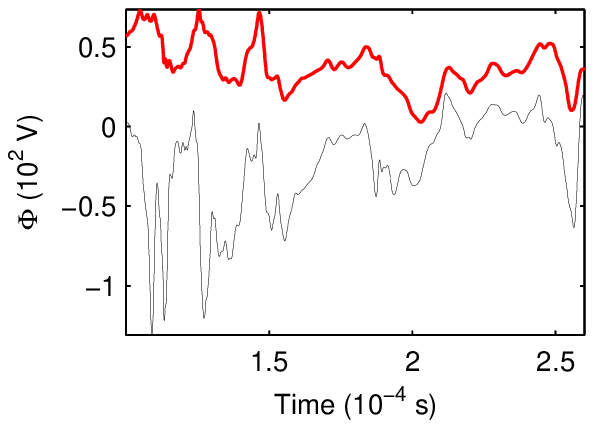}}
\caption{\sl (a) L-Mode time series of $\Phi$ (red bold line) and $V_{\rm{fl}}$
  (black thin line) for a radial position $\sim 15.2$ mm $\approx 17.2
  \rho_{\rm{s}}$ outside the separatrix. (b) Detailed evolution of a
  $\unit[160]{\mu s}$ time frame} 
\label{fig:lmode:phi}
\end{figure}

Fig.~\ref{fig:lmode:phi} shows a comparison of the plasma potential $\Phi$ and the
floating potential $V_{\rm{fl}}$ for a position within SOL, computed by means
of eq.~\ref{eq:phifloat}. A clear difference is evident, as is to be expected,
both in mean values and fluctuations. Comparing different radial positions,
the mean offset between $\Phi$ and $V_{\rm{fl}}$ is decreasing from the inside
to the outside. In terms of mathematics, the difference is caused by a product
of the electron temperature in energy units, divided by $e$, and a
dimensionless quantity, which depends on the temperature as well: 

\begin{equation}
\Delta^{\rm{real}} = \left( \frac{k_{\rm{B}} T_{\rm{e}}}{e} \right) \ln \left(
\frac{A_{\rm{e}}}{A_{\rm{i}}} \sqrt{\frac{T_{\rm{e}}}{T_{\rm{e}}+T_{\rm{i}}}}
\sqrt{\frac{m_{\rm{i}}}{2 \pi m_{\rm{e}}}} \right) = \left( \frac{k_{\rm{B}}
  T_{\rm{e}}}{e} \right) \Delta. 
\label{eq:deltaeq}
\end{equation}

\begin{figure}
 \centering
 \subfigure[]{\label{fig:lmode:delta:3}\includegraphics[width=7.5cm]{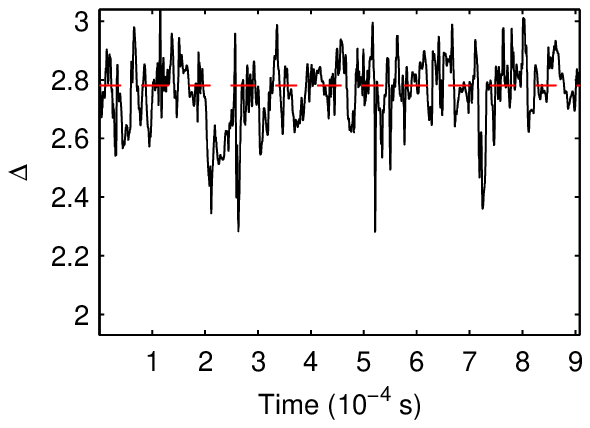}}\hfill
 \subfigure[]{\label{fig:lmode:deltareal:comp}\includegraphics[width=7.5cm]{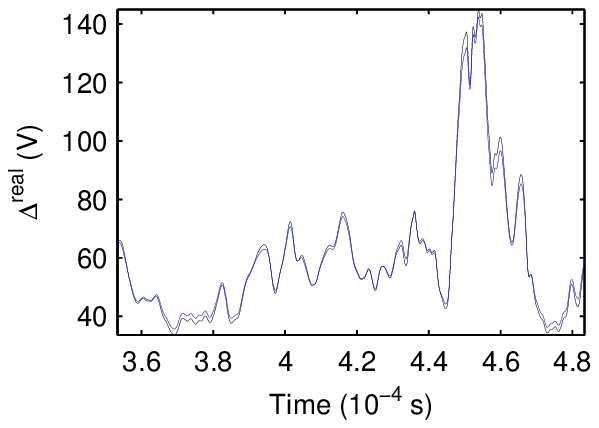}}
\caption{\sl (a) Normalised difference $\Delta$ between floating and plasma
  potential for a position $\sim 15.2$ mm $\approx 17.2 \rho_{\rm{s}}$ outside
  the separatrix, calculated from the fluctuating temperature time series
  (black solid line) and from constant temperature values averaged at the
  corresponding radial position (red dashed line). (b) Detailed evolution
  $\unit[130]{\mu s}$ of the actual difference $\Delta^{\rm{real}}$ between
  floating and plasma potential (black line) and $((k_{\rm{B}}
  T_{\rm{e}})/{e}) \Delta^{\rm{avg}}$ (blue line) within the SOL} 
\label{fig:lmode:delta}
\end{figure}

\begin{figure}
 \centering
 \subfigure[]{\label{fig:lmode:deltaprof:meandev}\includegraphics[width=7.5cm]{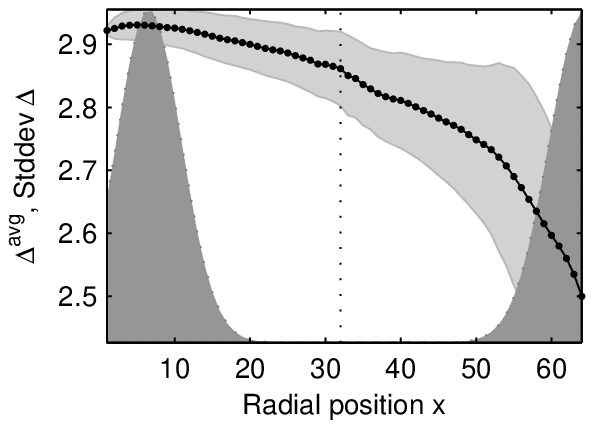}}\hfill  
 \subfigure[]{\label{fig:lmode:deltarealprof:meandev}\includegraphics[width=7.5cm]{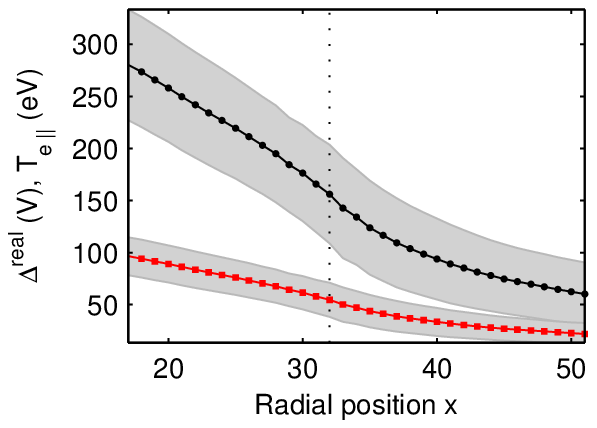}}
 \caption{\sl (a) Radial profile of $\Delta^{\rm{avg}}$ (black line with circles),
   standard deviation of exact $\Delta$ time series (light grey area) and
   source/sink regions of the \textit{GEMR} simulation model (dark grey
   shading). 
   (b) Radial profile of $\langle \Delta^{\rm{real}} \rangle$
   (black line with circles), $\langle T_{\rm{e}} \rangle$ (red line with
   squares) and the respective standard deviations (light grey area). The
   location of the separatrix is indicated by the dotted line} 
 \label{fig:lmode:deltaprof}
\end{figure}

If the temperature is assumed to be constant, i.e. all temperature
fluctuations are neglected, the latter can be understood as average difference
$\Delta^{\rm{avg}}$ between $\Phi$ and $V_{\rm{fl}}$ (also referred to as
$\alpha$), normalised by $e/(k_{\rm{B}} T_{\rm{e}})$ (see
\cite{Schrittwieser2002}). It can roughly be estimated for our nominal case
with $T_{\rm{i}} = 1.2 T_{\rm{e}}$ and $A_{\rm{i}} = A_{\rm{e}}$ to be
$\Delta^{\rm{avg}} \approx 2.79$. Thus on average the normalised difference
for a certain relation of $T_{\rm{e}}$ and $T_{\rm{i}}$ is constant and
deviations are mainly due to temperature fluctuations. On the other hand, the
background mid-pedestal temperature values used here are only reference
values. For all radial positions, but especially within the scrape-off layer
the simulated temperature relation can differ considerably from
this. Consequently, $\Delta^{\rm{avg}}$ is subject to radial variations (see
fig.~\ref{fig:lmode:deltaprof:meandev}). Taking into account that the temperature
is not constant results in a fluctuating time series $\Delta$, which is
plotted for one exemplary position within the SOL in
fig.~\ref{fig:lmode:delta:3}. The radial profile of its standard deviation is shown
as light grey area in fig.~\ref{fig:lmode:deltaprof:meandev}. In the same figure,
source and sink regions of the \textit{GEMR} simulation model are indicated by
dark grey shadings. The radial positions affected by these mechanisms are
removed in all subsequent profile plots, as they should be excluded from the
interpretation. 

It is important to note, that $\Delta$ is the \textit{normalised} difference
between $\Phi$ and $V_{\rm{fl}}$. It can be a useful quantity for rough
evaluations, if only average temperatures are available. A quantitative
estimation of the actual difference requires the calculation of $\Delta$
multiplied by $(k_{\rm{B}} T_{\rm{e}})/e$ (eq.~\ref{eq:deltaeq}) for every
time step and for all positions. Thus the shape and phase of the temperature
time series, which is varying depending on the measurement position, has a
direct influence on the difference between plasma and floating potential. 

As the mean temperature $\langle T_{\rm{e}} \rangle$ strongly decreases within
the radial simulation domain, the actual difference $\Delta^{\rm{real}}$
($\propto T_{\rm{e}} \Delta $) between floating and plasma potential is larger
inside the separatrix and becomes smaller in the SOL
(fig.~\ref{fig:lmode:deltarealprof:meandev}). The difference in terms of
fluctuation amplitudes seems to be somewhat larger inside and near the
separatrix, which is due to large temperature fluctuations in this region, but
a significant distinction of the general shape is evident at all radial
positions. 

In order to emphasise the role of the electron temperature fluctuations, a
comparison between the temporal evolution of $\Delta^{\rm{real}} =
((k_{\rm{B}} T_{\rm{e}})/{e}) \Delta$ and the quantity $((k_{\rm{B}}
T_{\rm{e}})/{e}) \Delta^{\rm{avg}}$ is shown in
fig.~\ref{fig:lmode:deltareal:comp}. There are only faint differences visible,
indicating the predominance of the fluctuations of $T_{\rm{e}}$ compared to
the fluctuations of $\Delta$ and, consequently, to the fluctuations of
$T_{\rm{i}}$. This behaviour can be observed in the entire radial domain, with
a slightly decreasing difference between the two terms plotted in
fig.~\ref{fig:lmode:deltareal:comp} from the inside to the outside and therefore a
decreasing importance of a precise $\Delta$. 

As stated in ref.~\cite{Adamek2002}, the differences in fluctuations of
$V_{\rm{fl}}$ and $\Phi$ are determined not only by fluctuations of the
electron temperature, but also by the phase relation between temperature and
potential. This becomes evident from the difference of the corresponding root
mean square values, 
\begin{equation}
[\rm{RMS}(\tilde{V}_{\rm{fl}})]^2 - [\rm{RMS}(\tilde{\Phi})]^2 =
[\rm{RMS}(\tilde{\Delta}^{\rm{real}})]^2 - 2 \langle \tilde{\Phi}
\tilde{\Delta}^{\rm{real}} \rangle. 
\label{eq:rmsdifference}
\end{equation}

\begin{figure}
 \centering
 \subfigure[]{\label{fig:lmode:xcorr:tephifluct:diffrms}\includegraphics[width=7.5cm]{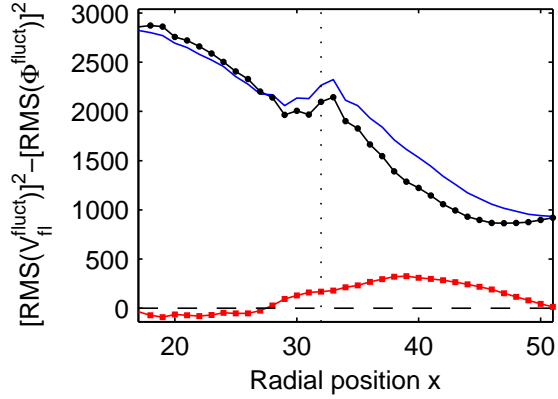}} 
\caption{\sl (a) Radial profile of $[\rm{RMS}(\tilde{V}_{\rm{fl}})]^2 -
  [\rm{RMS}(\tilde{\Phi})]^2$ (black line with circles), which is equal to the
  difference of $[\rm{RMS}(\tilde{\Delta}^{\rm{real}})]^2$ (blue solid line)
  and $2 \langle \tilde{\Phi} \tilde{\Delta}^{\rm{real}} \rangle$ (red line
  with squares) The tilde indicates fluctuating parts and RMS root mean square
  values} 
\label{fig:lmode:xcorr:tephifluct}
\end{figure}

 The tilde indicates fluctuating parts and the RMS of, for example,
$\tilde{\Phi}$ is given by $\langle \tilde{\Phi}^2 \rangle^{1/2}$, which in
the case of fluctuations with zero mean reflects the standard deviation
(except for a slightly different
normalisation). Fig.~\ref{fig:lmode:xcorr:tephifluct} shows radial profiles of the
left-hand side of eq.~\ref{eq:rmsdifference} and the two terms on the right-hand
side. It is apparent, that although the contribution of
$\tilde{\Delta}^{\rm{real}}$ ($\propto \tilde{T}_{\rm{e}}$) is clearly
dominant, the influence of the cross phase between $\tilde{\Phi}$ and
$\tilde{\Delta}^{\rm{real}}$, represented by $\langle \tilde{\Phi}
\tilde{\Delta}^{\rm{real}} \rangle$, is of non-negligible magnitude and
radially varying, with largest absolute values near the separatrix and within
the SOL, but rather small values inside the separatrix. A comprehensive
analysis of the phase relation between temperature and potential at different
radial positions by means of time-resolved Wavelet methods is currently in
progress and will be presented in a future work. 

The present calculation of the floating potential using simulated temperature
time series depends to a certain degree on the artificial offset factor
$\lambda_{\rm{os}}$, which has a decisive influence on the temperature
averages and therefore also on the mean offset between $\Phi$ and
$V_{\rm{fl}}$. Apart from that, time series of $\Delta$ are affected, as the
logarithm in the conversion equation eq.~\ref{eq:phifloat} reacts quite sensitive
on variations of the temperature minima, whose difference to zero is defined
by $\lambda_{\rm{os}}$. As a consequence, not only the distinct negative peaks
in fig.~\ref{fig:lmode:delta:3}, but also the radial profiles of
$\Delta^{\rm{avg}}$ and $\rm{Stddev} (\Delta)$ change their characteristics
depending on the chosen offset factor. Since $\lambda_{\rm{os}}$ has no effect
on temperature fluctuations, its impact on the fluctuating part and the
standard deviation of $\Delta^{\rm{real}}$ and on the comparison in
fig.~\ref{fig:lmode:deltareal:comp} is weak. That implies, that although the shape
of the $\Delta$ time series and the mean offset between $\Phi$ and
$V_{\rm{fl}}$ is affected by this numerical issue and the artificial offset,
the influence of electron and ion temperature fluctuations on calculations of
the floating potential can nevertheless be studied. This applies even more to
calculations of the radial particle flux, because only potential gradients are
of importance there. A prerequisite for any considerations about the influence
of small fluctuations on potential measurements is the question, whether an
experimental probe is in principle able to react almost instantaneously to
temperature fluctuations. This can indeed be assumed, because density
fluctuations, which have similar characteristics and a comparable ratio of
fluctuations to mean, can be detected.

\subsection{Radial particle flux}
\label{sec:lmode:particleflux}

One of the most important quantities for the statistical analysis of
experimental measurements is the turbulent fluctuation-induced averaged radial
particle flux (\cite{Carreras1996}--\cite{Endler1995} and
\cite{Schrittwieser2008}), defined by 
\begin{equation}
\Gamma_{\rm{r}} = \langle \tilde{n} \tilde{v_{\rm{r}}} \rangle \approx
\frac{\langle \tilde{n} \tilde{E}_{\rm{pol}}\rangle}{B}, 
\label{eq:flux}
\end{equation}
 where the fluctuating radial drift velocity is assumed to be the radial
 component of the fluctuating $\mathbf{\tilde{E}} \times \mathbf{B}$ velocity,
 $\tilde{v}_{\rm{r}} \approx v_{\mathbf{\tilde{E}} \times \mathbf{B}} =
 \tilde{E}_{\rm{pol}} / B$. Appropriate averaging is indicated by $\langle
 \cdot \rangle$. The poloidal electric field can be approximated by taking the
 difference between two potential measurements divided by their poloidal
 separation distance. In our simulations, this distance is given by $d_{12}
 \approx \unit[2.5]{mm}$. 
 Replacing the density by the expression for the ion saturation current
 $I_{\rm{is}}$ from eq.~\ref{eq:isat} yields 
\begin{equation}
\Gamma_{\rm{r}} \approx \frac{1}{B} \left\langle \left(\frac{1}{A_{\rm{i}} e}
\sqrt{\frac{m_{\rm{i}}}{k_{\rm{B}} (T_{\rm{e}}+T_{\rm{i}})}}
I_{\rm{is}}\right) \left(\frac{V_{\rm{fl},1}-V_{\rm{fl},2}}{d_{12}}\right)
\right\rangle. 
\label{eq:measuredfluxflux}
\end{equation}

\begin{figure}
 \centering
 \subfigure[]{\label{fig:lmode:flux:3}\includegraphics[width=7.5cm]{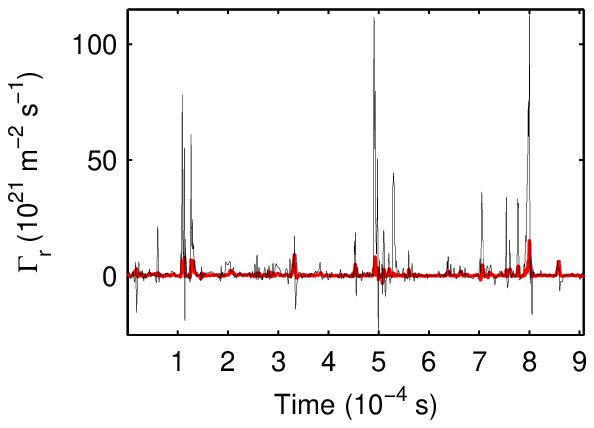}}\hfill
 \subfigure[]{\label{fig:lmode:flux:3part}\includegraphics[width=7.5cm]{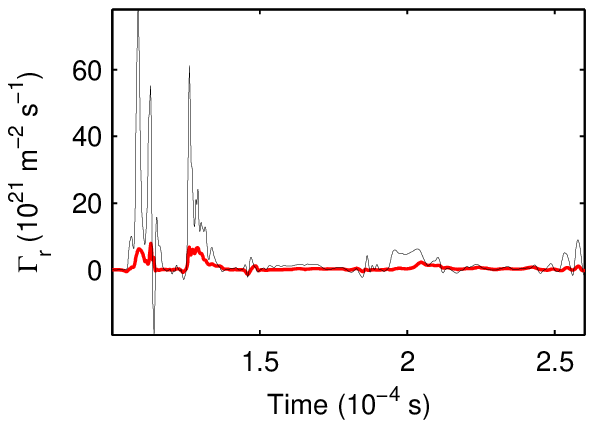}}
\caption{\sl Radial particle flux $\Gamma_{\rm{r}}$ calculated from plasma
  potential and density ($\Gamma_{\rm{r}}^{\rm{real}}$, red bold line) and
  from floating potential and ion saturation current
  ($\Gamma_{\rm{r}}^{\rm{m}}$, black thin line). The data is plotted for a
  radial position $\sim 15.2$ mm $\approx 17.2 \rho_{\rm{s}}$ outside the
  separatrix. (b) Detailed evolution of a $\unit[160]{\mu s}$ time frame.} 
\label{fig:lmode:flux}
\end{figure}

\begin{figure}
 \centering
 \subfigure[]{\label{fig:lmode:fluxprof:mean}\includegraphics[width=7.5cm]{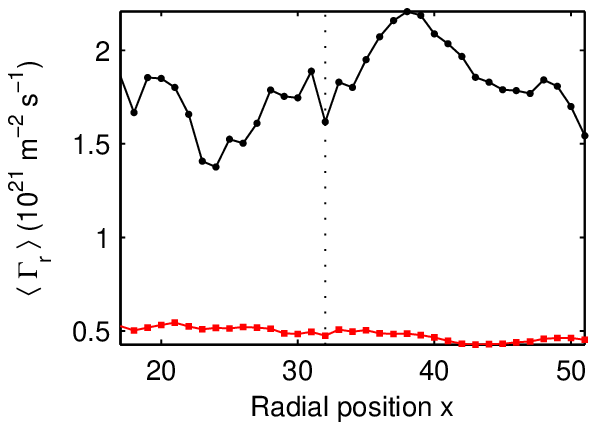}}\hfill
 \subfigure[]{\label{fig:lmode:fluxprof:dev}\includegraphics[width=7.5cm]{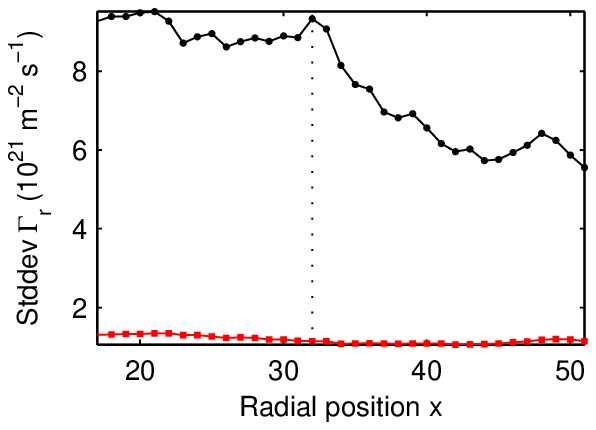}}
\caption{\sl (a) Radial profiles of the time averaged particle flux $\langle
  \Gamma_{\rm{r}} \rangle$ (averaged across the entire simulation time), (b)
  standard deviation $\rm{Stddev} (\Gamma_{\rm{r}})$. Real values are plotted
  as red line with squares, measured values as black line with circles} 
\label{fig:lmode:fluxprof}
\end{figure}

Both, the synthetically measured particle flux $\Gamma_{\rm{r}}^{\rm{m}}$
(using $V_{\rm{fl}}$ and densities inferred from $I_{\rm{is}}$ and averaged
temperatures) and the real particle flux $\Gamma_{\rm{r}}^{\rm{real}}$ (using
code outputs $\Phi$ and $n_{\rm{e}}$) have been calculated and their temporal
evolution, representing the instantaneous particle flux, is plotted in
fig.~\ref{fig:lmode:flux}. The time series of $\Gamma_{\rm{r}}^{\rm{m}}$ exhibit
much larger fluctuations than that of $\Gamma_{\rm{r}}^{\rm{real}}$ at all
radial positions, but especially inside and near the separatrix, which is
clearly visible in the radial profile of the standard deviation of the
time-averaged flux (fig.~\ref{fig:lmode:fluxprof:dev}). On average, $\rm{Stddev}
(\Gamma_{\rm{r}}^{\rm{m}})$ is $\sim 7$ times larger than $\rm{Stddev}
(\Gamma_{\rm{r}}^{\rm{real}})$. Magnitudes and radial profiles of the mean
fluxes $\langle \Gamma_{\rm{r}}^{\rm{real}} \rangle$ and $\langle
\Gamma_{\rm{r}}^{\rm{m}} \rangle$ depend on the length of the investigated
time range, but significant differences can be observed in any case
(fig.~\ref{fig:lmode:fluxprof:mean}, $\langle \Gamma_{\rm{r}}^{\rm{m}} \rangle \sim
3.5$ times larger than $\langle \Gamma_{\rm{r}}^{\rm{real}} \rangle$). The use
of an averaged temperature value in the density calculation only gives rise to
small deviations (see fig.~\ref{fig:lmode:isat}) and is not responsible for the
large discrepancy in terms of fluctuations. Hence it must primarily be due to
the radial velocity calculated from the gradient of the floating potential. 

The equivalent use of the floating potential instead of the plasma potential,
which is the usual procedure to estimate the particle flux from Langmuir probe
measurements, is based on the assumption, that the difference between plasma
and floating potential is roughly constant at two closely spaced locations. As
only the difference between two adjacent time series is relevant, the error
caused by using $V_{\rm{fl}}$ instead of $\Phi$ is expected to be small,
depending on different deviations from this constant at the two
locations. Such deviations have been discussed in the previous section, but
for radially varying positions, whereas here different positions in the
$y$-direction of the simulation model are of importance. Fluctuations of time
series at these positions correspond to fluctuations of two poloidally
separated positions, because $k_{\parallel} \ll k_{\perp}$. 
For $V_{\rm{fl}} = \Phi - [(k_{\rm{B}} T_{\rm{e}})/e] \Delta$, the difference
between approximations of plasma potential gradient and floating potential
gradient is given by 
\begin{eqnarray}
\label{eq:diffgradphi}
\frac{\Phi_{1}-\Phi_{2}}{d_{12}} - \frac{V_{\rm{fl},1}-V_{\rm{fl},2}}{d_{12}}
& = & \frac{1}{d_{12}} \left[\Phi_{1}-\Phi_{2}- \left(\Phi_{1}-\frac{k_{\rm{B}}
    T_{\rm{e},1}}{e} \Delta_{1}\right) + \left(\Phi_{2}-\frac{k_{\rm{B}}
    T_{\rm{e},2}}{e} \Delta_{2}\right)\right] \nonumber \\ 
 & = & \frac{1}{d_{12}} \frac{k_{\rm{B}}}{e} \left(T_{\rm{e},1} \Delta_{1} - T_{\rm{e},2} \Delta_{2}\right).
\end{eqnarray}

\begin{figure}
 \centering
 \subfigure[]{\label{fig:lmode:dphij:3}\includegraphics[width=7.5cm]{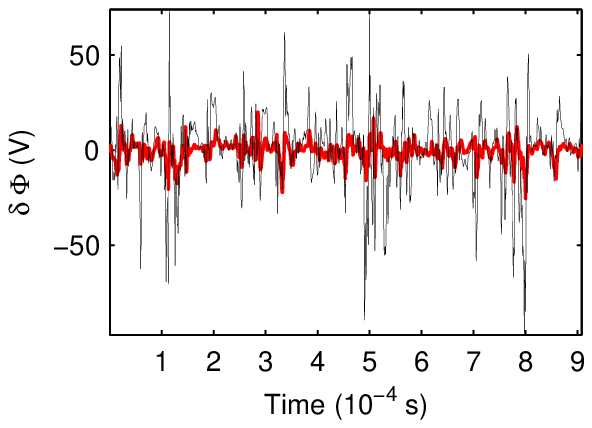}}\hfill
 \subfigure[]{\label{fig:lmode:dphij:3part}\includegraphics[width=7.5cm]{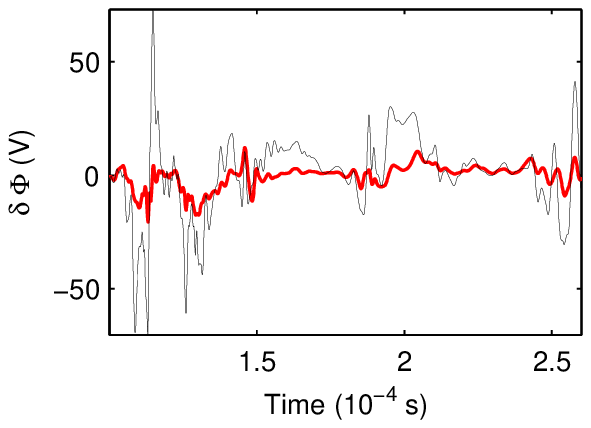}}
\caption{\sl (a) Difference between two time series $\Phi_{1}$ and $\Phi_{2}$ (red
  bold line) and between $V_{\rm{fl},1}$ and $V_{\rm{fl},2}$ (black thin
  line), collected at the poloidally adjacent positions used for the
  estimation of $\tilde{v_{\rm{r}}}$. The data is plotted for a radial
  position $\sim 15.2$ mm $\approx 17.2 \rho_{\rm{s}}$ outside the
  separatrix. (b) Detailed evolution of a $\unit[160]{\mu s}$ time frame.} 
\label{fig:lmode:dphij} 
\end{figure}

 This expression vanishes, if $T_{\rm{e},1} = T_{\rm{e},2}$ and $\Delta_{1} =
 \Delta_{2}$ is assumed for the two adjacent positions. According to
 fig.~\ref{fig:lmode:dphij}, which shows $\delta \Phi = \Phi_{1}-\Phi_{2}$ and
 $\delta V_{\rm{fl}} = V_{\rm{fl},1}-V_{\rm{fl},2}$, this is clearly not the
 case, so there must be a significant difference between these quantities. The
 second part of equation eq.~\ref{eq:diffgradphi} can be further expanded by
 splitting the respective quantities into mean values and fluctuations: 
\begin{eqnarray}
\label{eq:diffgradphifluct}
\left(T_{\rm{e},1} \Delta_{1} - T_{\rm{e},2} \Delta_{2}\right) = &
\Big[ \big( \langle T_{\rm{e},1} \rangle \langle \Delta_{1} \rangle - \langle
  T_{\rm{e},2} \rangle \langle \Delta_{2} \rangle \big) + \big(\langle
  T_{\rm{e},1} \rangle \tilde{\Delta}_{1} - \langle T_{\rm{e},2} \rangle
  \tilde{\Delta}_{2} \big) \nonumber\\ 
& + \big(\langle \Delta_{1} \rangle \tilde{T}_{\rm{e},1} - \langle \Delta_{2}
  \rangle \tilde{T}_{\rm{e},2} \big) + \big(\tilde{T}_{\rm{e},1}
  \tilde{\Delta}_{1} - \tilde{T}_{\rm{e},2} \tilde{\Delta}_{2} \big) \Big].
\end{eqnarray}

\begin{figure}
 \centering
 \subfigure[]{\label{fig:lmode:dtfluct:tjmp1}\includegraphics[width=5.2cm]{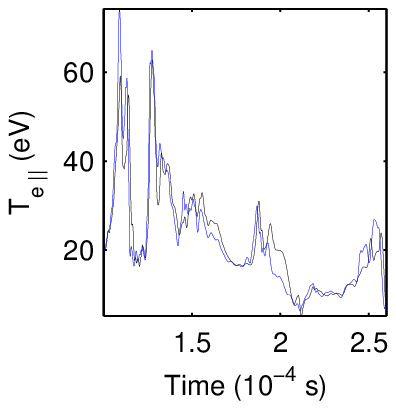}}
 \subfigure[]{\label{fig:lmode:dtfluct:dtfluctjmp1}\includegraphics[width=5.2cm]{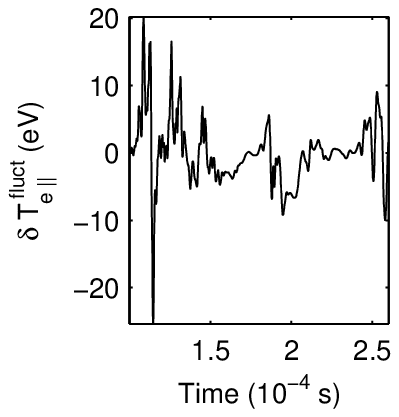}}
 \subfigure[]{\label{fig:lmode:dtfluct:dphi3smalldtfluct}\includegraphics[width=5.2cm]{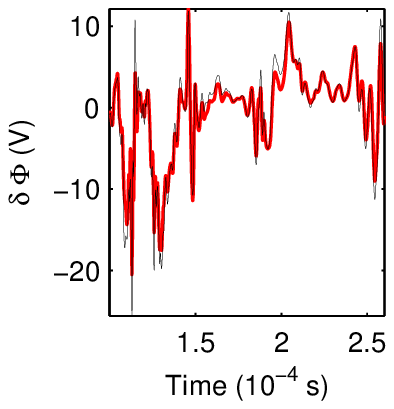}}  
\caption{\sl (a) Temperature time series $T_{\rm{e},1}$ (black line) and
  $T_{\rm{e},1}$ (blue line), collected at the two poloidally adjacent
  positions. (b) Difference between the fluctuating parts of the two
  temperature time series. (c) Difference between the two spatially adjacent
  time series for $\Phi$ (red bold line) and $V_{\rm{fl}}$ (black thin line)
  for the artificial case of small differences of the fluctuating temperature
  time series (see text). All plots show the detailed evolution of a
  $\unit[160]{\mu s}$ time frame for a radial position $\sim 15.2$ mm $\approx
  17.2 \rho_{\rm{s}}$ outside the separatrix} 
\label{fig:lmode:dtfluct} 
\end{figure}

 A small offset between the two time series of $\delta \Phi$ and $\delta
 V_{\rm{fl}}$ is caused by slightly different mean values of $T_{\rm{e},1}$
 and $T_{\rm{e},2}$ as well as $\Delta_{1}$ and $\Delta_{2}$
 (fig.~\ref{fig:lmode:dtfluct:tjmp1}), which is expressed by the first term on the
 right-hand side of eq.~\ref{eq:diffgradphifluct}. The remaining terms, the second
 last of which providing by far the largest contribution, are related to the
 strong fluctuations of $\delta V_{\rm{fl}}$. These originate from differences
 in the fluctuating parts of $T_{\rm{e},1}$ and $T_{\rm{e},2}$
 (fig.~\ref{fig:lmode:dtfluct:dtfluctjmp1}) and, accordingly but of less
 importance, $\Delta_{1}$ and $\Delta_{2}$. 

To point this out, $V_{\rm{fl}}$ and the differences $\delta \Phi$ and $\delta
V_{\rm{fl}}$ of spatial adjacent time series have been calculated for the
artificial case of temperature time series, which have equal mean and
therefore no offset at the two positions, and whose difference in fluctuations
is the same as before (fig.~\ref{fig:lmode:dtfluct:dtfluctjmp1}), but with much
smaller amplitude ($1/10$). The result is plotted in
fig.~\ref{fig:lmode:dtfluct:dphi3smalldtfluct}, where $\delta V_{\rm{fl}}$ is in
substantial better agreement with $\delta \Phi$ than in
fig.~\ref{fig:lmode:dphij:3}. 

\begin{figure}
 \centering
 \subfigure[Inside
   separatrix]{\label{fig:lmode:fluxpdf:1}\includegraphics[width=5.2cm]{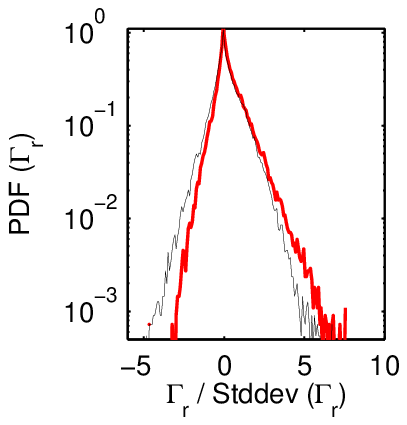}}\hfill 
 \subfigure[Near
   separatrix]{\label{fig:lmode:fluxpdf:2}\includegraphics[width=5.2cm]{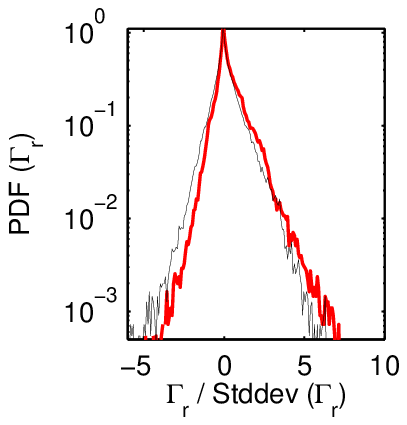}}\hfill 
 \subfigure[Outside
   separatrix]{\label{fig:lmode:fluxpdf:3}\includegraphics[width=5.2cm]{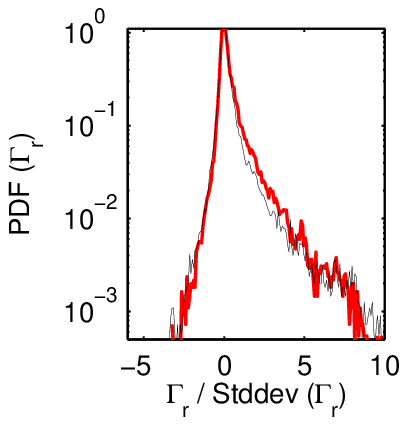}} 
\caption{\sl PDF of the radial particle flux $\Gamma_{\rm{r}}$, calculated from
  plasma potential and density (red bold line) and from floating potential and
  ion saturation current (black thin line). The data is plotted at the radial
  positions (a) inside, (b) near and (c) outside the separatrix as specified
  in fig.~\ref{fig:lmode:raw}} 
\label{fig:lmode:fluxpdf}
\end{figure}

\begin{figure}
 \centering
 \subfigure[]{\label{fig:lmode:momentprof:skewx}\includegraphics[width=7.5cm]{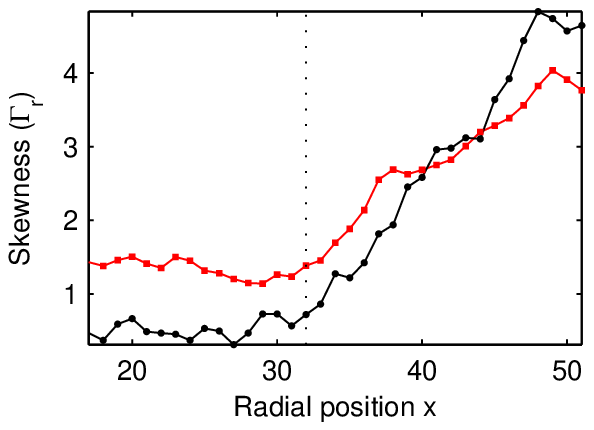}}
 \subfigure[]{\label{fig:lmode:momentprof:kurtx}\includegraphics[width=7.5cm]{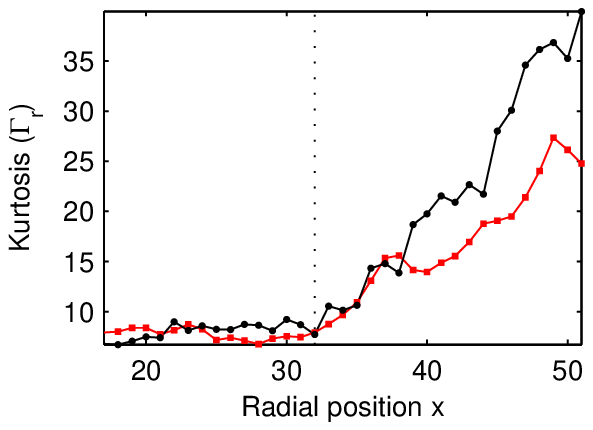}}
\caption{\sl Radial profiles of (a) the Skewness and (b) the Kurtosis of real (red
  line with squares) and measured particle flux $\Gamma_{\rm{r}}$ (black line
  with circles)} 
\label{fig:lmode:momentprof}
\end{figure}

In order to investigate the impact of temperature fluctuations on statistical
properties, probability distribution functions (PDF) have been computed. As
shown in fig.~\ref{fig:lmode:fluxpdf} and fig.~\ref{fig:lmode:momentprof}, there is a
different behaviour inside and outside the SOL. Inside and near the
separatrix, the PDF of the real particle flux features a larger skewness at
most radial positions, whereas the kurtosis is largely of about the same
magnitude. In the scrape-off layer, from a radial position of $\sim
\unit[6]{mm} \approx 6.8 \rho_{\rm{s}}$ outside the separatrix, both the
skewness and the kurtosis of $\Gamma_{\rm{r}}^{\rm{real}}$ are smaller than
that of $\Gamma_{\rm{r}}^{\rm{m}}$, which indicates, that in the time series
of $\Gamma_{\rm{r}}^{\rm{m}}$ strong deviations from the mean value occur even
more frequently than in $\Gamma_{\rm{r}}^{\rm{real}}$. 

Above investigations show, that both flux time series and their statistics are
differing to a significant degree depending on whether temperature
fluctuations are taken into account for the calculations. This seems to be the
case not only for measurement positions inside and near the separatrix, but
also within the SOL, where Langmuir probe measurements are typically
located. Although the quantitative results of our numerical simulation might
differ from the experimental situation in a fusion device and should not be
taken as a reference, the importance of the fluctuating electron temperature
is nevertheless evident. It should be mentioned, that variations of the offset
factor $\lambda_{\rm{os}}$, which is used to compensate numerical errors of
the simulated time series, only have small influence on this and do not change
the overall flux features.

\section{Ideal ballooning mode blowout}

\begin{figure}
 \centering
 \subfigure[]{\label{fig:hmode:raw:tepl}\includegraphics[width=7.5cm]{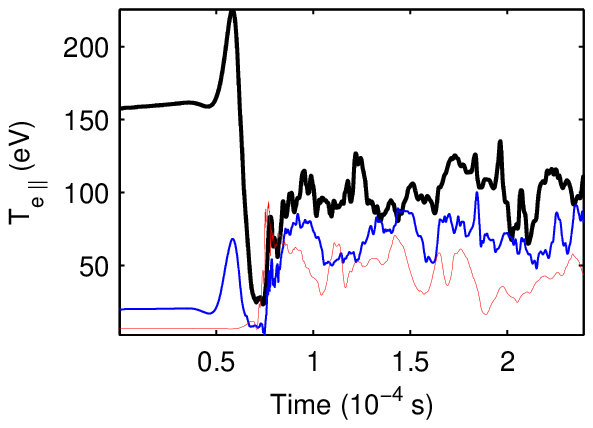}}\hfill
 \subfigure[]{\label{fig:hmode:raw:all}\includegraphics[width=7.5cm]{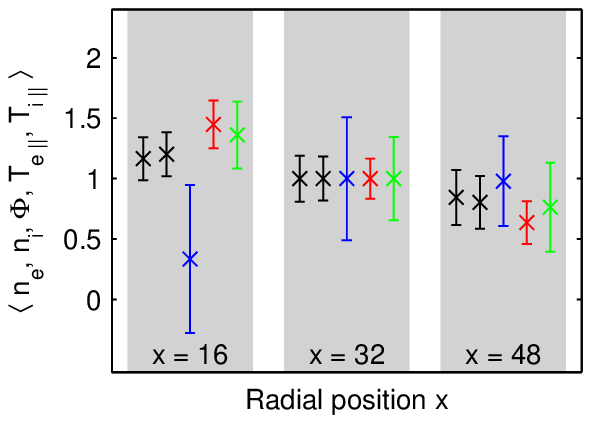}}
\caption{\sl (a) Time series of the electron temperature $T_{\rm{e}
    \parallel}$. (b) Mean values and standard deviations of electron and ion
  density $n_{\rm{e}}$ and $n_{\rm{i}}$ (black), plasma potential $\Phi$
  (blue), electron temperature $T_{\rm{e} \parallel}$ (red) and ion
  temperature $T_{\rm{i} \parallel}$ (green), normalised to the respective
  mean at $x = 32$, in the aftermath of an IBM blowout at different radial
  positions. The simulation grid point $x = 16$ (black bold line in (a))
  corresponds to the radial distance of $\sim -14.8$ mm $\approx -11.8
  \rho_{\rm{s}}$, $x = 32$ (blue line in (a)) to $\sim 0.4$ mm $\approx 0.34
  \rho_{\rm{s}}$ and $x = 48$ (red thin line in (a)) to $\sim 15.2$ mm
  $\approx 12.2 \rho_{\rm{s}}$, measured from the separatrix} 
\label{fig:hmode:raw}
\end{figure}

As a next step, a simulated ELM type-I like ideal ballooning mode (IBM)
situation has been analysed, which exhibits a large interchange blowout. A
sudden burst connected with enhanced fluctuations is clearly visible in the
time series of $T_{\rm{e} \parallel}$, which is shown as an example in
fig.~\ref{fig:hmode:raw:tepl}, but of course also in the signals of $n_{\rm{e}}$
and $n_{\rm{i}}$, $\Phi$, and $T_{\rm{i} \parallel}$. In the aftermath of the
blowout the standard deviations of the temperatures and densities are lower
compared to the L-mode case at most radial positions, but the standard
deviations of the potential time series are larger
(fig.~\ref{fig:hmode:raw:all}). The background temperature and density parameters
are $T_{\rm{e}} = \unit[300]{eV}$, $T_{\rm{i}} = \unit[360]{eV}$ and
$n_{\rm{e}} = n_{\rm{i}} = 2.5 \cdot 10^{19}$ m$^{-3}$. Gradient lengths and
background magnetic field remain unchanged compared to the L-mode case of the
previous section. This yields $\beta_{\rm{e}} \approx 4.0 \cdot 10^{-4}$ and
allows the occurrence of an IBM instability, according to the MHD ideal
ballooning criterion $\alpha_{\rm{m}} = q^{2} R \nabla \beta > \hat{s}$ (with
$\beta = 2 \beta_{\rm{e}}$, safety factor $q = 1.5 + 3.5 (r/a)^{2}$ and
magnetic shear parameter $\hat{s} = (r / q) (\partial q / \partial r)
\rightarrow \hat{s}_{a} = 1.4$ at the separatrix). Heat and density sources
are set to a small value, in order to provide an IBM blowout as pure as
possible and to prevent distortions caused by other effects. The averaged grid
size perpendicular to the magnetic field is $0.75 \rho_{\rm{s}} \times 0.98
\rho_{\rm{s}}$ (with $\rho_{\rm{s}} \approx 1.25$ mm). The analysed IBM time
series have a length of $15000$ data points and a temporal resolution of $\sim
\unit[0.0167]{\mu s}$ (total duration $\unit[0.25]{ms}$). All signals of this
section are acquired near the outboard midplane. Further details on IBM
simulations with \textit{GEMR} can be found in ref.~\cite{Kendl2010}.  

\subsection{Plasma density vs. ion saturation current and plasma potential
  vs. floating potential}

\begin{figure}
 \centering
 \subfigure[]{\label{fig:hmode:isat:3}\includegraphics[width=7.5cm]{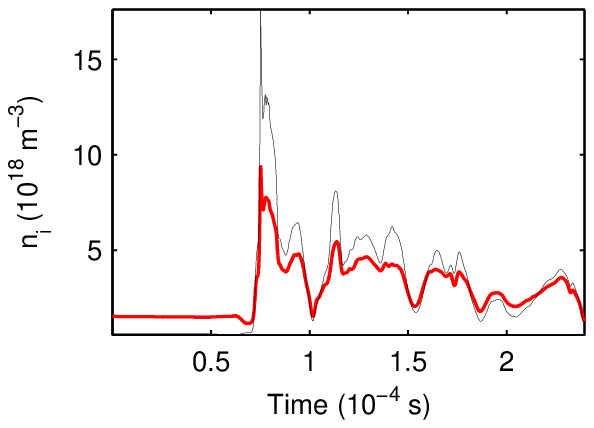}}\hfill
 \subfigure[]{\label{fig:hmode:phi:3}\includegraphics[width=7.5cm]{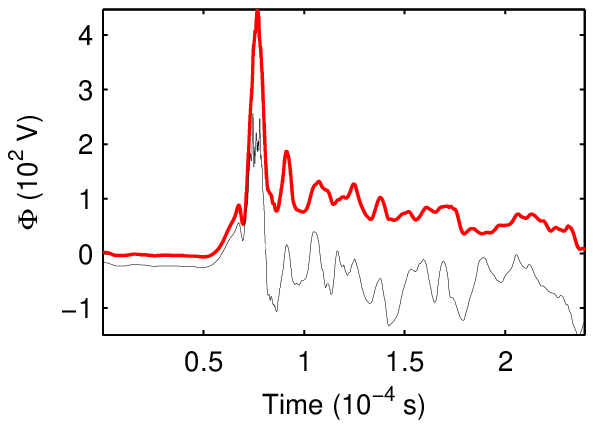}}
\caption{\sl(a) IBM time series of code output $n_{\rm{i}}$ (red bold line) and $n_{\rm{i}}^{\rm{avg}}$ (black thin line) calculated from $I_{\rm{is}}$ using averaged values for $T_{\rm{e} \parallel}$ and $T_{\rm{i} \parallel}$. (b) IBM time series of $\Phi$ (red bold line) and $V_{\rm{fl}}$ (black thin line). The data is plotted for a radial position $\sim 15.2$ mm $\approx 12.2 \rho_{\rm{s}}$ outside the separatrix.}
\label{fig:hmode:isatphi}
\end{figure}

Just as in the L-Mode case discussed above, there are no major differences
between the time series of density and ion saturation current and both feature
similar characteristics. If the density is inferred from $I_{\rm{is}}$ by
means of averaged temperature values (fig.~\ref{fig:hmode:isat:3}), the main
features are preserved, apart from the time-dependent offset mentioned in
subsection \ref{sec:lmode:nvsisat}, which results from averaging. As the use
of time-averaged values is more defective for large density and temperature
variations, it is most pronounced during the blowout. 

\begin{figure}
 \centering
 \subfigure[]{\label{fig:hmode:deltareal:13}\includegraphics[width=7.5cm]{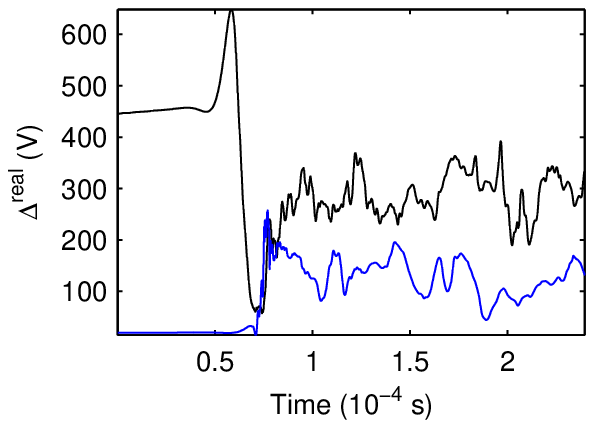}}\hfill
 \subfigure[]{\label{fig:hmode:deltarealprof:mean}\includegraphics[width=7.5cm]{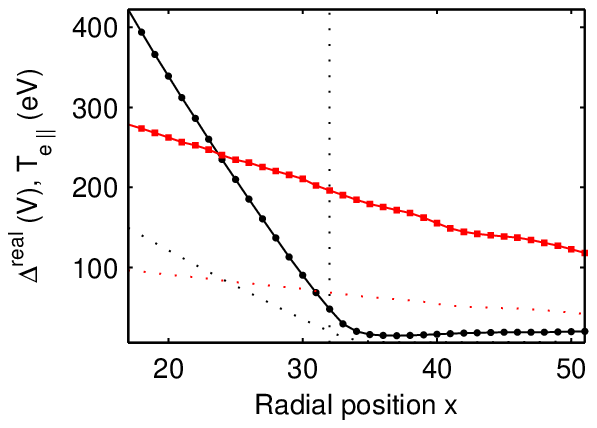}}
\caption{\sl (a) Actual difference $\Delta^{\rm{real}}$ between floating and
  plasma potential for the radial positions $\sim 14.8$ mm $\approx 11.8
  \rho_{\rm{s}}$ inside (black line) and $\sim 15.2$ mm $\approx 12.2
  \rho_{\rm{s}}$ outside the separatrix (blue line). (b) Radial profile of
  $\langle \Delta^{\rm{real}} \rangle$ before (black line with circles) and
  after the blowout (red line with squares). Profiles of $\langle T_{\rm{e}}
  \rangle$ are plotted as dotted line.} 
\label{fig:hmode:deltareal}
\end{figure}

$\Phi$, $V_{\rm{fl}}$ and $\Delta^{\rm{real}}$, the actual difference between
plasma and floating potential, have also been calculated for the IBM case
(fig.~\ref{fig:hmode:phi:3} and fig.~\ref{fig:hmode:deltareal:13}). An individual
investigation of $\Delta$ has been omitted, since according to
section~\ref{sec:lmode:phivsvfl} the inclusion of the factor $T_{\rm{e}}$
plays an important role. The predominance of $\tilde{T}_{\rm{e}}$ compared to
$\tilde{\Delta}$ is still present, as a comparison between
$\Delta^{\rm{real}}$ and $((k_{\rm{B}} T_{\rm{e}})/{e}) \Delta^{\rm{avg}}$,
shown in fig.~\ref{fig:lmode:deltareal:comp} for the L-mode situation, yields
similar results for IBM time series.  
Differences between fluctuations of $\Phi$ and $V_{\rm{fl}}$ are clearly
noticeable at all radial positions, but due to the limited lengths of the IBM
time series no reliable results for the standard deviation could be obtained,
in particular before and during the blowout. After the blowout, the profile of
the standard deviation seems to be similar to the L-mode case. 
Inside the separatrix, the IBM blowout leads to a strong reduction of the
temperature, whereas near and beyond the separatrix it is increased. This is
reflected in the temporal evolution of
$\Delta^{\rm{real}}$. Fig.~\ref{fig:hmode:deltarealprof:mean} shows radial profiles
of $\langle \Delta^{\rm{real}} \rangle$ and $\langle T_{\rm{e}} \rangle$,
separately averaged across the time range immediately before and after the
blowout. Inside the separatrix the time average of the actual difference and
the temperature is considerably larger before the blowout than
afterwards. From a point near but not across the separatrix it is just the
other way around. The decrease or increase of $\Delta^{\rm{real}}$ after the
blowout has implications if the plasma potential is calculated from
$V_{\rm{fl}}$ using temperatures averaged across the entire time domain, as
the resulting time series will be either underestimated before the blowout and
overestimated afterwards or vice-versa. 

\subsection{Radial particle flux}

\begin{figure}
 \centering
 \subfigure[]{\label{fig:hmode:fluxprof:ibm:mean}\includegraphics[width=7.5cm]{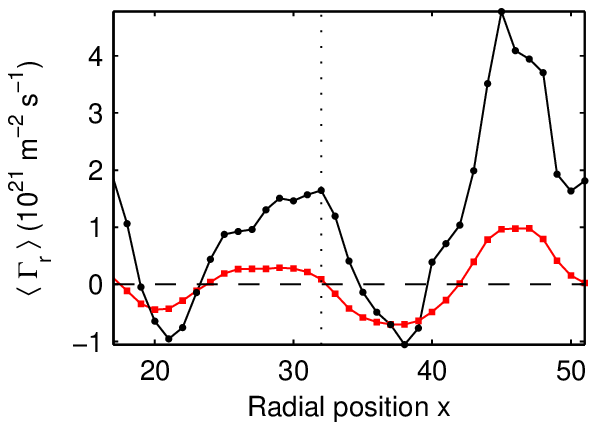}}\hfill
 \subfigure[]{\label{fig:hmode:fluxprof:ibm:dev}\includegraphics[width=7.5cm]{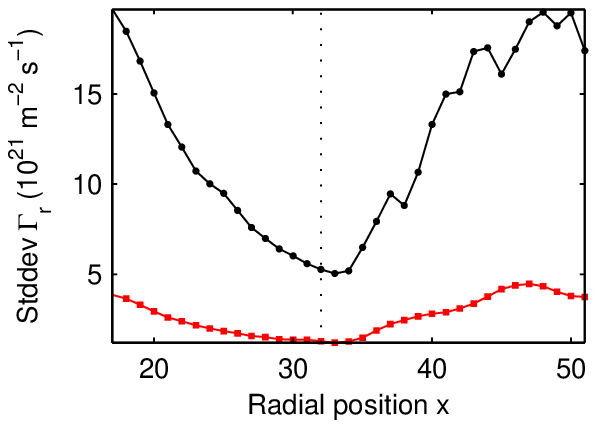}}
\caption{\sl (a) Radial profiles of the time averaged particle flux $\langle
  \Gamma_{\rm{r}} \rangle$, (b) standard deviation $\rm{Stddev}
  (\Gamma_{\rm{r}})$ during the IBM blowout ($\unit[29.1]{\mu s} --
  \unit[97.3]{\mu s}$). Real values are plotted as red line with squares,
  measured values as black line with circles} 
\label{fig:hmode:fluxprof:ibm}
\end{figure}

Also in the IBM case, the instantaneous "measured" radial particle flux
$\Gamma_{\rm{r}}^{\rm{m}}$ calculated from $I_{\rm{is}}$ and $V_{\rm{fl}}$
shows much larger fluctuations than the real flux
$\Gamma_{\rm{r}}^{\rm{real}}$. This is again primarily due to the use of
$V_{\rm{fl}}$ instead of $\Phi$. The averaged temperature in the calculation
of density from $I_{\rm{is}}$ leads to a further amplification of the
fluctuation amplitudes. The magnitude of strong flux peaks during the blowout,
which varies considerably at different radial positions, has a decisive impact
on radial profiles of mean fluxes and standard deviations, if these are
averaged across the entire simulation time. Hence the time spans during and
after the IBM blowout should be treated individually. In
fig.~\ref{fig:hmode:fluxprof:ibm}, the time range $\unit[29.1]{\mu s} -
\unit[97.3]{\mu s}$ has been chosen, which includes the large peaks of the
blowout at all radial positions. Fig.~\ref{fig:hmode:fluxprof:ibm} shows profiles
from the time range $\unit[97.3]{\mu s} - \unit[239.5]{\mu s}$. Small
displacements of the transition point yield slightly different profiles, but
the general characteristics remain unchanged. 

\begin{figure}
 \centering
 \subfigure[]{\label{fig:hmode:fluxprof:aibm:mean}\includegraphics[width=7.5cm]{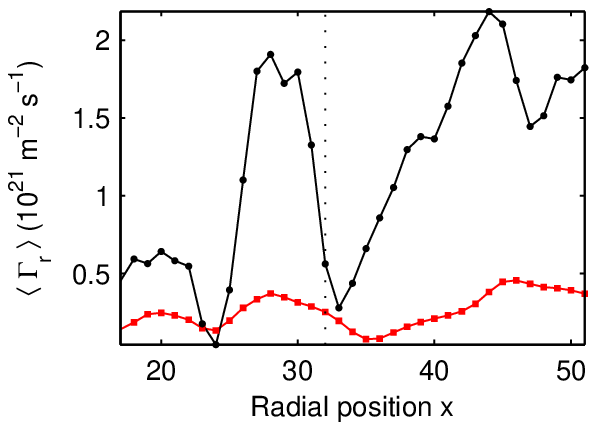}}
 \subfigure[]{\label{fig:hmode:fluxprof:aibm:dev}\includegraphics[width=7.5cm]{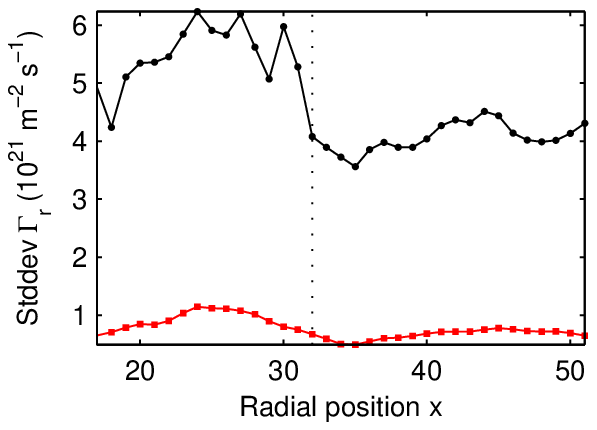}}
\caption{\sl (a) Radial profiles of the time averaged particle flux $\langle
  \Gamma_{\rm{r}} \rangle$, (b) standard deviation $\rm{Stddev}
  (\Gamma_{\rm{r}})$ after the IBM blowout ($\unit[97.3]{\mu s} --
  \unit[239.5]{\mu s}$). Real values are plotted as red line with squares,
  measured values as black line with circles} 
\label{fig:hmode:fluxprof:aibm}
\end{figure}

The larger amplitudes of $\Gamma_{\rm{r}}^{\rm{m}}$ compared to
$\Gamma_{\rm{r}}^{\rm{real}}$ can be observed at most positions in the radial
profiles of mean values and standard deviations. The ratios between measured
and real values are approximately comparable to the values given in the L-mode
section. As shown in fig.~\ref{fig:hmode:fluxprof:ibm:mean}, $\langle
\Gamma_{\rm{r}}^{\rm{m}} \rangle$ and $\langle \Gamma_{\rm{r}}^{\rm{real}}
\rangle$ differ at some radial positions not only in absolute values but also
in sign, which is a significant distortion of the actual situation. However,
the respective mean values are based only on a relatively small time interval
and should not be overemphasised. Interestingly, the fluctuation amplitudes
and the standard deviation of $\Gamma_{\rm{r}}^{\rm{m}}$ and
$\Gamma_{\rm{r}}^{\rm{real}}$ approach a local minimum near the
separatrix. This is especially pronounced during the blowout
(fig.~\ref{fig:hmode:fluxprof:ibm:dev}), but is still present afterwards
(fig.~\ref{fig:hmode:fluxprof:aibm:dev}). 

As our simulation code can only provide data for the temporal evolution of one
single IBM blowout and its aftermath, the fluctuations covered by the time
series are not sufficient to allow a more detailed statistical analysis
including probability density functions and statistical moments.

\section{Conclusions}

From the above investigations can be concluded, that neglecting any
temperature fluctuations in our computations of probe measurements results in
significant differences between the synthetically measured values and the
actual quantities at all radial positions of the simulation domain, both in a
saturated L-mode situation and in simulations of an IBM blowout. This holds
especially for the floating potential and its use in calculations of the
fluctuation induced particle flux. The latter shows a considerable discrepancy
not only in the temporal evolution but also in statistical properties, if
spatial variations of the temperature fluctuations are not taken into
account. However, fluctuations of the electron temperature seem to be of much
greater importance than ion temperature fluctuations. 

Compared to the saturated L-mode situation, the investigation of time series involving an IBM
blowout did not yield a major alteration of the ratio between real and
measured quantities, apart from unsurprising changes due to the large peaks. 

Although a realistic implementation of a virtual Langmuir probe in
terms of geometry and particle processes would require the use of a kinetic
model, the results from our gyrofluid simulations can nevertheless be regarded
as relevant for experimental measurements, as all comparisons are performed
within the simulation. In this respect they may serve as an indication to the
particular importance of electron temperature fluctuations, whose neglect in
evaluations of experimental data might lead to substantial deviations from the
actual quantities. 

\newpage

\section*{Acknowledgements}
We would like to thank B. D. Scott, author of the {\it GEMR} gyrofluid
model, R. Schrittwieser, F. Mehlmann and J. Peer for their assistance and
valuable discussions. This work was supported by the Austrian Science Fund FWF
under Contract No. Y398, by a junior research grant
({\textquotedblleft}Nachwuchsf\"{o}rderung{\textquotedblright}) from
University of Innsbruck, and by the European Commission under the 
Contract of Association between EURATOM and \"{O}AW, carried out within the 
framework of the European Fusion Development Agreement (EFDA). The views and
opinions expressed herein do not necessarily reflect those of the European
Commission. Furthermore, this work was also supported by the Austrian Ministry
of Science BMWF as part of the UniInfrastrukturprogramm of the
Forschungsplattform Scientific Computing at LFU Innsbruck. 


\end{document}